\documentclass[traditabstract]{aa}
\usepackage[utf8]{inputenc}
\usepackage{amsmath}
\usepackage{natbib}
\usepackage{url}
\usepackage{graphicx}
\usepackage{epsfig}
\usepackage{txfonts}
\usepackage{color}
\usepackage{multirow}
\usepackage{amssymb}
\usepackage{epsfig}
\bibpunct{(}{)}{;}{a}{}{,} 
\usepackage[plainpages=false]{hyperref}

\title{Is there a highly magnetized neutron star in \hbox{GX~301$-$2}?}
\author{V. Doroshenko\inst{1}, A. Santangelo\inst{1}, V. Suleimanov\inst{1,5},
 I. Kreykenbohm\inst{2,3}, R. Staubert\inst{1}, C. Ferrigno\inst{1,4}, D. Klochkov\inst{1}}
\institute{
Institut für Astronomie und Astrophysik, Sand 1, 72076 Tübingen, Germany \and
Dr. Karl Remeis-Sternwarte, Sternwartstrasse 7, 96049 Bamberg, Germany \and
Erlangen Centre for Astroparticle Physics (ECAP), Erwin-Rommel-Strasse 1, 91058 
Erlangen, Germany\and
ISDC Data Centre for Astrophysics, Chemin d'Écogia 16, CH-1290 Versoix, Switzerland\and
Kazan State University, Kremlevskaya 18, 420008, Kazan, Russia
}

\begin{document}

\bibliographystyle{aa}

\abstract{
We present the results of an in-depth study of the long-period X-ray pulsar
\hbox{GX~301$-$2}. Using archival data of \textsl{INTEGRAL}, \textsl{RXTE
ASM}, and \textsl{CGRO BATSE}, we study the spectral and timing properties
of the source. Comparison of our timing results with previously published
work reveals a secular decay of the orbital period at a rate of
$\simeq-3.25\times10^{-5}\,\mathrm{d\,yr}^{-1}$, which is an order of
magnitude faster than for other known systems. We argue that this is
probably result either of the apsidal motion or of gravitational coupling
of the matter lost by the optical companion with the neutron star, although
current observations do not allow us to distinguish between those
possibilities. We also propose a model to explain the observed long pulse
period. We find that a very strong magnetic field $B\sim10^{14}$\,G can
explain the observed pulse period in the framework of existing models for
torques affecting the neutron star. We show that the apparent contradiction
with the magnetic field strength $B_\mathrm{CRSF}\sim4\times10^{12}$\,G
derived from the observed cyclotron line position may be resolved if the
line formation region resides in a tall accretion column of
height~$\sim$2.5--3\,$R_{\mathrm{NS}}$. The color temperature measured from
the spectrum suggests that such a column may indeed be present, and our
estimates show that its height is sufficient to explain the observed
cyclotron line position. 
}

\authorrunning{V. Doroshenko et al.}
\maketitle

\section{Introduction}
\label{sec:introduction} 
GX~301$-$2 (also known as 4U~1223$-$62) is a high-mass \hbox{X-ray} binary
system, consisting of a neutron star orbiting the early \hbox{B-type}
optical companion Wray~977. The neutron star is a $\sim$680\,s X-ray pulsar
\citep{White:1976p993}, accreting from the dense wind of the optical
companion. The wind's mass-loss rate of the optical component is one of the
highest known in the galaxy: \hbox{$\dot{M}_\mathrm{loss}\sim10^{-5}
M_{\odot}\,\mathrm{yr}^{-1}$} \citep{Kaper:2006p1357}. Because the terminal
velocity of the wind is very low
\hbox{($w_{0}\sim300-400~\mathrm{km}\,\mathrm{s}^{-1}$},\,\citealt{Kaper:20
06p1357}), the accretion rate is high enough to explain the observed
luminosity of $L_{\mathrm{X}}\sim10^{37} \mathrm{erg}\,\mathrm{s}^{-1}$.
The distance to the source is estimated to be between
\hbox{$1.8\pm0.4$\,kpc} \citep{Parkes:1980p1360} and
5.3\,kpc~\hbox{\citep{Kaper:1995p151}}, depending on the spectral
classification of Wray~977. The latest estimate is
3\,kpc~\citep{Kaper:2006p1357}. The orbit is highly eccentric with an
eccentricity of \ensuremath{\sim}0.5 and an orbital period of
\ensuremath{\sim41.5\,d}~\citep{Koh:1997p138}. The absence of X-ray
eclipses despite the large radius ($R\sim43$\,R$_{\odot}$) of Wray~977
\citep{Parkes:1980p1360} constrains the inclination angle in the range
$44-78^\circ$ with a best-fit value of
$i\sim66^\circ$~\citep{Kaper:2006p1357, Leahy:2008p358}. The source
exhibits regular X-ray flares about 1--2\,d before the periastron passage
(orbital phase $\sim0.95$). There is also an indication of a second flare
at orbital phase $\sim0.5$~\citep{Koh:1997p138}. Several hypotheses have
been proposed to explain the observed orbital lightcurve, including a
circumstellar disk~\citep{Koh:1997p138} and a quasi-stable accretion
stream~\citep{Leahy:2008p358}. Similar to other wind accreting systems, the
pulse period behavior of \hbox{GX~301$-$2} on short time scales is
described well by a random walk model~\citep{deKool:1993p147}.
\hbox{GX~301$-$2} exhibits a long-term pulse period evolution as well. The
observed pulse period remained \ensuremath{\sim}700\,s until 1984 when it
began to decrease during a rapid spin-up episode observed by \textsl{BATSE}
\citep{Koh:1997p138,Bildsten:1997p2328}. The spin-up trend reversed in 1993
\citep{Pravdo:2001p360} and ever since the pulse period has been increasing
\citep{LaBarbera:2005p156,Kreykenbohm:2004p155}. 

The X-ray spectrum of the \hbox{GX~301$-$2} is rich in features. The lower
energy range is subject to heavy and variable photoelectric absorption
\citep{White:1976p993}. As shown by \cite{Kreykenbohm:2004p155} and
\cite{LaBarbera:2005p156}, a partial covering model with two absorption
columns is required to describe the spectrum. There is a complex of iron
lines at $\sim$6.4 to 7.1\,keV~\citep{Watanabe:2003p357}. A high-energy
cutoff at \hbox{$\sim20$\,keV}, together with a deep and broad cyclotron
resonance scattering feature (CRSF) at $\sim$30--45\,keV, is present at
higher
energies~\citep{Makishima:1992p3220,Orlandini:2000p153,Kreykenbohm:2004p155
, LaBarbera:2005p156}. The CRSF is highly variable with pulse phase, and it
exhibits interesting correlations with the continuum
parameters~\citep{Kreykenbohm:2004p155}. 

The nature of accreting pulsars with long pulse periods is still poorly
understood. Because of the low moment of inertia of the neutron star, the
accelerating torque of the accreted matter can spin up a neutron star very
efficiently. Braking torques are then required to explain the observed long
pulse periods. It is commonly assumed that the observed pulse period is
determined by the equality of torques affecting the neutron star or relaxes
to the value determined by this equality. Braking torques are generally
associated with the coupling of the neutron star's magnetic field with the
surrounding plasma. The drag force depends on the relative linear speed of
field lines at certain effective radius, which in turn depends on the
magnetic field strength. The efficiency of braking decreases for slowly
rotating and weakly magnetized neutron stars so a strong field (up to
$10^{15}$\,G, \citealt{Shakura:1975p2764}) is required to spin down a
slowly rotating accreting X-ray pulsar even further. This results in an
apparent contradiction with field estimates obtained from the CRSF centroid
energy, which is
$B\sim(E_\mathrm{cyc}\mathrm{keV}/11.57)\times10^{12}\mathrm{G}\sim4\times10^{12}\mathrm{G}$ 
in the case of \hbox{GX~301$-$2} and in the same order of magnitude as for
other sources. 

We suggest that this contradiction may be resolved if the line-forming
region resides in an accretion column of significant height
\citep{Basko:1976p1538}, comparable to the neutron star radius. We
investigate this hypothesis using \textsl{INTEGRAL} and \textsl{BATSE}
observations to study the spectral and timing properties of
\hbox{GX~301$-$2}. 

\section{Observations and data selection}
The International Gamma-Ray Astronomy Laboratory (\textsl{INTEGRAL})
launched in October 2002 by the European Space Agency (ESA) is equipped
with 3 co-aligned coded mask instruments: \textsl{IBIS} (Imager onboard the
\textsl{INTEGRAL} Satellite,~\citealt{Ubertini:2003p1120}), \textsl{JEM-X}
(Joint European X-ray Monitor,~\citealt{Lund:2003p1129}), and \textsl{SPI}
(Spectrometer on \textsl{INTEGRAL},~\citealt{Vedrenne:2003p1138}). Because
of limited \textsl{SPI} sensitivity for variable sources, we rely on data
from \textsl{IBIS} (the \textsl{ISGRI} layer) and \textsl{JEM-X} in this
paper. Among the \textsl{INTEGRAL} instruments, \textsl{IBIS} has the
largest field of view and, therefore, the highest probability of observing
the source. We used a total of 554 available public pointings with
\hbox{GX~301$-$2} within the \textsl{IBIS} half-coded field of view for the
pulse period determination (i.e. for Table~\ref{tab:perhist}). These data
include a long observation that covers $\sim60$\% of the orbital cycle and
is long enough to allow binary ephemeris estimation (283 pointings in
\textsl{INTEGRAL} revolutions 322-330). Three dedicated observations (see
Table~\ref{tab:obs}) were also performed during the pre-periastron flare
and were used to study the spectrum of the source. 

We also used results provided by the \textsl{ASM/RXTE} teams at
\textsl{MIT} and at the \textsl{RXTE SOF} and \textsl{GOF} at
\textsl{NASA}'s \textsl{GSFC} and \textsl{CGRO BATSE} pulsar
\textsl{DISCLA} histories data by \cite{Bildsten:1997p2328} to study the
long-term evolution of the spin period.

\section{Observational results}
\label{sec:data_analysis}
\subsection{Timing analysis}
\label{sub:timing_analysis}
To derive the intrinsic pulse period of the source, the lightcurve must be
corrected for Doppler delays due to the orbital motion of the source and
the satellite. Phase connection or pulse time arrival analysis is a precise
timing technique, based on measuring arrival times of individual pulses or
groups of pulses~\citep{staubert2009}. It allows to determine the Doppler
delays and therefore the orbital parameters of the system. A fixed phase of
the pulsating flux from a pulsar is observed at times 
\citep{Nagase:1982p2720}:
\begin{equation}
\begin{array}{l}
 \displaystyle T_n=T_0+P_0n+\frac{1}{2}\dot{P}P_0n^2 +\frac{1}{6}\ddot{P}{P_0}^2n^3 +...\\
 \displaystyle \qquad {} ...+ a\sin(i)F_n(e,\omega,T_\mathrm{PA},\theta)
\end{array}
\label{eq:tn}
\end{equation}
referred to as Time Of Arrival (TOA), where ${P}_{0}$, $\dot{P}$, and
$\ddot{P}$ are the intrinsic pulse period and its time derivatives at the
initial epoch $ {T}_{0}$. The last term represents the Doppler delays due
to the orbital motion as a function of the Kepler parameters for an
eccentric orbit: the projected semi-major axis $ a\sin{i}$ in light seconds
($ i$ is the orbit inclination), the eccentricity $e$, the longitude of the
periastron $ \omega $, time of periastron passage $ T_\mathrm{PA} $, and
the mean anomaly $\theta =2\pi{(T-T_\mathrm{PA})}/{P_\mathrm{orb}}$. To
obtain a solution for the unknown pulse and orbital parameters, a number of
measurements of $ {T}_{n}$ (for known $n$) must be obtained. Usually only $
{T}_{n,obs}$ is measured, while $ n$ must be found during the fitting
procedure to obtain a self-consistent solution. The orbital period may be
estimated either directly as one of the free parameters or by comparing
periastron passage times of subsequent cycles (i.e. similarly to the pulse
period). The latter method is more precise (see e.g.
\citealt{staubert2009}). 

\begin{figure}[t]
	\centering
		\includegraphics{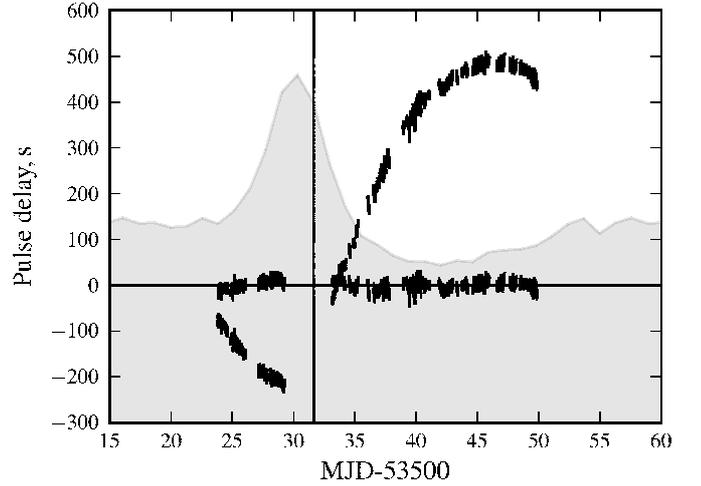}
 \caption{Time delays of pulse arrival times induced by the orbital motion.
Changes due to the intrinsic variation of the spin period are subtracted.
best-fit residuals are also shown. The best-fit periastron passage time is
marked with a vertical line. The folded \textsl{RXTE ASM} orbital profile
with the pre-periastron flare is plotted in gray.} 
	\label{fig:toa}
\end{figure}

Using archival \textsl{ISGRI} observations and the standard \textsl{OSA}
6.1 software provided by \textsl{ISDC}\footnote{http://isdc.unige.ch}, we
constructed lightcurves with 20\,s time bins in the energy range
20--40\,keV and determined the pulse arrival times (each pulse profile
obtained by folding $\sim20$ individual pulses) for data from revolutions
322-330 using a technique similar to the one by~\cite{Koh:1997p138}. This
is the only \textsl{INTEGRAL} observation to cover a significant fraction
of the orbital cycle, and it allows estimation of binary parameters. We
then used Eq.~\ref{eq:tn} to determine $P$,$\dot{P}$ and $T_\mathrm{PA}$.
The other orbital parameters were fixed to values reported
by~\cite{Koh:1997p138}. Our best-fit values are
$P_\mathrm{pulse}=684.1618(3)\,\mathrm{s}$,
$\dot{P}_\mathrm{pulse}=4.25(22)\times10^{-8}\,\mathrm{s}\,\mathrm{s}^{-1}$
at the epoch $53523.8$, and $T_\mathrm{PA}=53531.63(1)$\,MJD. All
uncertainties are at $1\sigma$ confidence level unless otherwise stated.
Pulse delays from the orbital motion and residuals of the best-fit are
plotted in Fig.~\ref{fig:toa}. 
Comparing our $T_\mathrm{PA}$ value with the historical values reported by
\cite{ws84}, \cite{Sato:1986p130} and~\cite{Koh:1997p138} allows estimation of the orbital period.
\begin{figure}[t]
 \centering
 \includegraphics{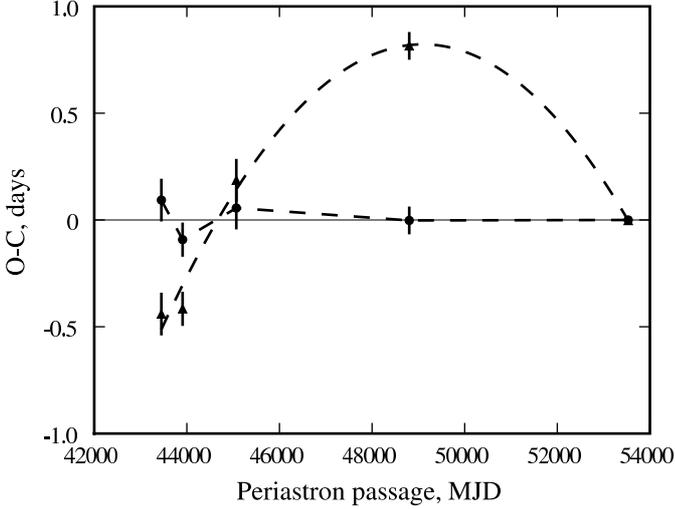}
 \caption{Residuals to fit over periastron passage times for orbital period
with (circles) and without (triangles) inclusion of the orbital period
derivative.} 
 \label{fig:orbper}
\end{figure}
Inclusion of our measurement requires introducing of a secular change to
the orbital period. The quality of the fit improves
significantly\footnote{The $\chi^2$ drops from 52 to 0.52 with an F-test
significance of $\sim98\%$}. The residuals to fit both with and without
inclusion of a secular change are plotted in Fig.~\ref{fig:orbper}. Our
best fit values are $P_\mathrm{orb}=41.506\pm0.003$\,d and
$\dot{P}_\mathrm{orb}=(-3.7\pm0.5)\times10^{-6}\,\mathrm{s}\,\mathrm{s}^{-1
}$ at the reference time reported by~\cite{Sato:1986p130}:
$T_\mathrm{PA,0}=43906.06\pm0.11$. This estimate is consistent with the
direct measurements of the orbital period both by~\cite{Sato:1986p130} and
by~\cite{Koh:1997p138}. 

It should be emphasized that the commonly used value of
P$_\mathrm{orb}$=41.498\,d by~\cite{Koh:1997p138} was obtained by
comparison of the $T_\mathrm{PA}$ values as well (the authors compared
their value to that by~\cite{Sato:1986p130} under the assumption of a
constant orbital period). On the other hand, all published measurements
including ours are consistently described when an orbital period derivative
is included. For the time of the \textsl{INTEGRAL} observation, the
predicted orbital period \hbox{is $\sim41.472$\,d.} The periastron passage
time measured with the orbital period value fixed to this prediction does
not change significantly: $T_\mathrm{PA}=53531.65\pm0.01$\,MJD. 

Because the pre-periastron flare in the orbital lightcurve of the source is
associated with the periastron passage time, an additional check can be
made using the long-term lightcurve of the source. We split a 10-year long
\textsl{RXTE ASM} barycentered daily lightcurve of \hbox{GX~301$-$2} (all
bands combined) into parts of $\sim5$ orbital cycles in length and folded
each part with the orbital period $P_\mathrm{orb}$=41.498\,d to obtain a
series of orbital profiles. The relative phase shifts and the associated
orbital periods were then determined in the same way as for the pulse
period. The best-fit value for a constant period is
$P_\mathrm{orb}=41.482\pm0.001$\,d. The mean value of the orbital period,
calculated using $P_\mathrm{orb}$ and $\dot{P}_\mathrm{orb}$ obtained
above, is consistent with the observed value at the time of the
\textsl{ASM} observations, although an orbital period derivative is not
formally required by the \textsl{ASM} data alone. 

\begin{figure}[t]
 \resizebox{\hsize}{!}{\includegraphics{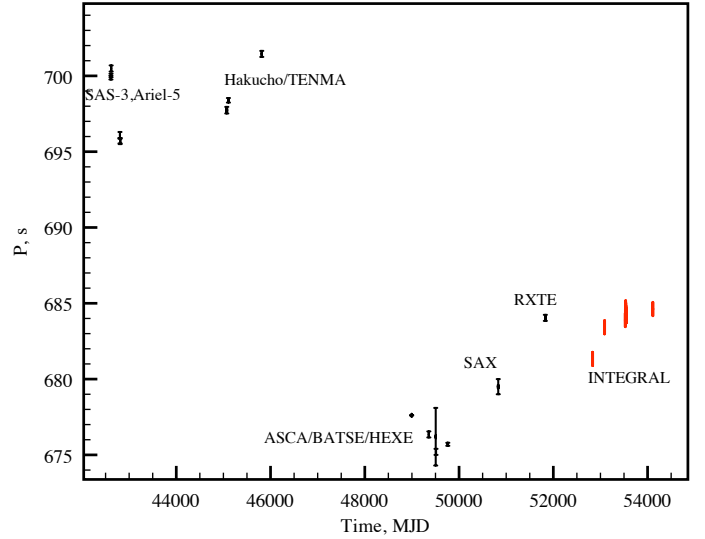}}
\caption{Long-term pulse period evolution of \hbox{GX~301$-$2}.}
\label{fig:perhist}
\end{figure}

A set of pulse period measurements with epoch folding was performed with
the updated ephemeris. We grouped all available \textsl{INTEGRAL} data by
the observation time by the ``k-means'' clustering
algorithm~\citep{MacQueen:1966p2718}. The number of groups was chosen such
that the mean number of X-ray pulses within one group was $\sim50$. We
searched for pulsations in each of the groups with epoch
folding~\citep{Larsson:1996p368}. A few groups where no pulsations were
found because of insufficient statistics were rejected afterwards. The
results are listed in Table~\ref{tab:perhist} and plotted in
Fig.~\ref{fig:perhist} together with historical values for clarity. On
average, \hbox{GX~301$-$2} continues to spindown after the first 
\textsl{INTEGRAL} observation of the source. 
\begin{table}
\centering \caption{Pulse period values obtained with archival \textsl{INTEGRAL} data.}
\label{tab:perhist}
\begin{tabular}{llll}
\hline
\hline
{MJD$_{\text{obs}}$} & {Period,\,s.} & {MJD$_{\text{obs}}$} & {Period,\,s.}\\
\hline
 52833.5 & $681.33\pm0.04$ & 53538.0 & $684.59\pm0.12$\\
 53088.0 & $683.42\pm0.07$ & 53541.6 & $684.4\pm0.04$\\
 53525.9 & $683.9\pm0.1$ & 53545.6 & $684.27\pm0.04$\\
 53528.5 & $684.19\pm0.16$ & 53549.4 & $684.15\pm0.05$\\
 53529.7 & $684.15\pm0.44$ & 54111.9 & $684.62\pm0.06$\\
 53535.0 & $684.73\pm0.05$ & 54277.6 & $685.15\pm0.07$\\
\hline
\end{tabular}
\tablefoot{
An updated ephemeris was used to correct the lightcurve for binary motion.
The error is estimated as 10\% of the width of the peak in the periodogram.
}
\end{table}
\subsection{Spectral analysis} 
\label{sub:spectral_analysis}
The observations listed in Table~\ref{tab:obs} were used to obtain the
broadband spectrum of the source. Since all three observations were made at
almost the same orbital phase, we combined all data to have better
statistics. We used the standard OSA 6.1 pipeline for spectral extraction.
A systematic error of 1\% for all \textsl{ISGRI} and of 2\% for all
\textsl{JEM-X} spectra was assigned as suggested in the OSA documentation. 
\begin{table}
\caption{Pointed observations of \hbox{GX~301$-$2} by \textsl{INTEGRAL}, with
an updated ephemeris to calculate the orbital phase.
\label{tab:obs}}
\centering
\begin{tabular}{lllll}
\hline
\hline
\textsl{INTEGRAL} & MJD of & Orbital & Exposure, & Rate,\tablefootmark{*}\\
 science window & observation & phase & ksec & cts\,s$^{-1}$\\
\hline
 05180027--66 & 54110.5--12.2 & 0.96--1.02 & 91.66 & 98\\
 05730048--60 & 54276.3--76.9 & 0.95--0.97 & 31.98 & 163\\
 05740012--40 & 54277.6--78.9 & 0.99--1.02 & 69.38 & 76\\
\hline
\end{tabular}
\tablefoot{
\tablefoottext{*}{in 20-60\,keV energy range using the \textsl{ISGRI} lightcurve.}
}
\end{table}

Spectra of X-ray pulsars are usually described with phenomenological
multi-component models. The continuum of \hbox{GX~301$-$2} was been modeled
with a cut-off power law modified at low energy by photoelectric
absorption. An iron emission K$_\alpha$ line was been also observed. In
fact, there is a complex of iron lines at
$\sim6.4\sim7.1$\,keV~\citep{Watanabe:2003p357,LaBarbera:2005p156} in the
spectrum of \hbox{GX~301$-$2}. These are not resolved with \textsl{JEM-X}.
We therefore used a simple Gaussian-shaped profile with larger width to
formally describe this feature. 

The photoelectric absorption of the source's spectrum is strongly variable,
and at least two absorption columns are identified. Part of the X-ray
emission is thought to be strongly absorbed close to the neutron star,
while all emission from this region is also subject to absorption in the
overall stellar wind of the optical companion. A model describing this
physical situation is the absorbed partial covering
model~\citep{Kreykenbohm:2004p155,LaBarbera:2005p156}. 
 
From a more physical point of view, the spectrum of an accreting pulsar is
believed to be mainly the result of a Comptonization processes of thermal
photons in the accretion column and in the neutron star atmosphere. The
emerging spectrum depends on the optical depth and generally has a
power-law shape, with a cut-off at an energy corresponding to the
temperature of the Comptonizing medium
\citep[$\sim3kT_e$,][]{Sunyaev:1980p2243}. Phenomenological models aim at
describing this shape regardless of the optical depth. For
\hbox{GX~301$-$2} two models have been used in literature.
\citet{LaBarbera:2005p156} adopted a modified ``high energy'' cut-off,
while~\citet{Kreykenbohm:2004p155} used a so-called Fermi-Dirac cut-off. As
discussed in~\cite{pos}, both models describe the \textsl{INTEGRAL} data
well with parameters close to the published ones. It is somewhat difficult,
however, to interpret these results from a physical point of view. We
therefore focus here on a different description. 

One of the first physical models to describe Comptonization spectra was
proposed by~\cite{Sunyaev:1980p2243}. Compton scattering in strong magnetic
field is a more complicated
problem~\citep{Lyubarskii:1986p3028,Meszaros1985}, but for the saturated
case ($\tau_e\gg1$) a blackbody-like spectrum is formed in both
cases~\citep{Lyubarskii:1986p3028}. The~\hbox{\cite{Sunyaev:1980p2243}
}model is included in the standard \textsl{XSPEC} distribution as
\textsl{COMPST}. Free parameters include the electron temperature of the
medium $T_e$, optical depth $\tau_{e}$, and normalization $A_{st}$. We used
this model because it contains the least number of free parameters and
produces identical results to more complex models for \hbox{GX~301$-$2}.
The pulse-phase averaged spectrum was extracted and fitted with the
partially absorbed \textsl{COMPST} model. The fit results are listed in
Table~\ref{tab:spepars}. Since the optical depth of the Comptonizing medium
is very high, we verified that a simple black body model provides an
equally good description of the data. The unabsorbed source flux in the
same energy range is
\hbox{$\sim1.8\times10^{-8}$\,erg\,cm$^{-2}$\,s$^{-1}$} in both models. 

A CRSF was necessary in the fit. This was included assuming a
Gaussian-shaped profile. With the inclusion of the line the $\chi^2_{red}$
dropped from $\sim3.8$ (depending on the model) to values around 1.2 (see
Table~\ref{tab:spepars}). 
\begin{table}
\caption{Fit results for phase averaged spectra. Uncertainties
are expressed at 90\% confidence level.}
\label{tab:spepars} 
\centering
\begin{tabular}{llllllll}
\hline
\hline
                              Parameter &                  Absorbed &                  Absorbed\\
                                        &                 blackbody &           \textsl{COMPST}\\
\hline
$N_{H,1}[10^{22} \rm{atoms}/\rm{cm}^2]$ &                    $\le4$ &                    $\le4$\\
$N_{H,2}[10^{22} \rm{atoms}/\rm{cm}^2]$ &     $178.3_{-6.7}^{+6.9}$ &    $175.6_{-9.9}^{+10.4}$\\
                            $c_{\rm F}$ & $0.798_{-0.008}^{+0.008}$ &    $0.78_{-0.01}^{+0.01}$\\
            $E_\mathrm{gabs}[\rm{keV}]$ &      $45.8_{-1.6}^{+1.7}$ &      $45.9_{-1.6}^{+1.8}$\\
       $\sigma_\mathrm{gabs}[\rm{keV}]$ &      $15.0_{-1.7}^{+1.8}$ &      $15.1_{-1.8}^{+2.0}$\\
                      $d_\mathrm{gabs}$ & $57.45_{-17.89}^{+23.66}$ & $57.93_{-18.19}^{+25.93}$\\
                            $A_{bb/st}$ &    $0.31_{-0.04}^{+0.06}$ &    $0.35_{-0.05}^{+0.09}$\\
                         $T_e$\,[{keV}] &       $5.1_{-0.2}^{+0.2}$ &       $5.1_{-0.2}^{+0.3}$\\
                             $\tau_{e}$ &                        -- &      $42.3_{-3.9}^{+4.8}$\\
              $E_\mathrm{Fe}[\rm{keV}]$ &    $6.32_{-0.02}^{+0.02}$ &    $6.32_{-0.02}^{+0.02}$\\
         $\sigma_\mathrm{Fe}[\rm{keV}]$ &    $0.36_{-0.05}^{+0.04}$ &    $0.38_{-0.04}^{+0.04}$\\
\hline
                           $\chi^2$/dof &                  1.18/149 &                  1.17/148\\

\end{tabular}
\end{table}
\begin{figure}[t]
\centering
\resizebox{\hsize}{!}{\includegraphics{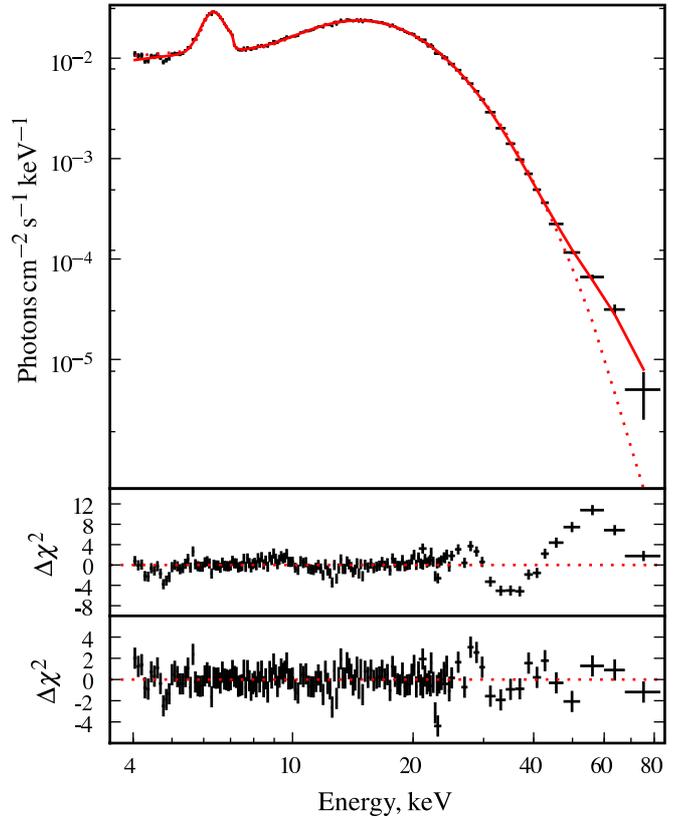}}
\caption{Unfolded spectrum and residuals for the \textsl{COMPST} model
without (dotted line and second panel) and with the inclusion (solid line
and bottom pane) of a CRSF at $\sim46$\,keV, using \textsl{ISGRI}
(20-80\,keV) and \textsl{JEM-X} (4-20\,keV) data.} 
\label{fig:phav}
\end{figure} 
\section{Discussion} 
\subsection{Orbital period evolution}
Our estimate of the rate of orbital period decay 
$\dot{P}_\mathrm{orb}/P_\mathrm{orb}\simeq-3.25\times10^{-5}\,\mathrm{yr}^{
-1}$ exceeds that of other known sources at least by one order of magnitude.
Previous detections include Cen~X$-$3
($\dot{P}_\mathrm{orb}/P_\mathrm{orb}=-1.738\times10^{-6}\,\mathrm{yr}^{-1}
$ \citep[and references therein]{bagot}, SMC~X$-$1
($\dot{P}_\mathrm{orb}/P_\mathrm{orb}=-3.36\times10^{-6}\,\mathrm{yr}^{-1}$) 
and Her~X$-$1 with
$\dot{P}_\mathrm{orb}/P_\mathrm{orb}=-1.0\times10^{-8}\,\mathrm{yr}^{-1}$~
\citep{staubert2009}. GX~301$-$2 is very different from all these systems.
It is younger and has a highly eccentric orbit, while other systems have
almost circular orbits. Both real and apparent changes in the orbital
period are expected to be greater for an eccentric orbit. 

We measured the rate of decay of the orbit by comparing the times of
several periastron-passages. Those are determined by fitting the observed
pulse delays as a function of orbital phase, and they may in principle be
correlated with other model parameters, particularly with the longitude of
periastron due to apsidal motion. The span of our data used for the pulse
time arrival analysis does not allow both $T_{PA}$ and $\omega$ to be
reliably constrained simultaneously. All published estimates of $\omega$
are also consistent 
with each other within
uncertainties, but still we cannot rule out that apsidal motion contributes to
the observed apparent change in the orbital period.

The eccentric orbit and very strong mass transfer in the system (the mass
loss by the optical component is \hbox{$\dot{M}_\mathrm{loss}\sim10^{-5}
M_{\odot}\,\mathrm{yr}^{-1}$}, \citealp{Kaper:2006p1357}) suggest that some
intrinsic changes in the orbital period are also expected. 

The optical companion is much heavier than the neutron star and contributes
almost nothing to the orbital angular momentum of the system. Direct mass
loss via the stellar wind by the optical companion therefore does not lead
to significant loss of the angular momentum. The optical star becomes less
massive, leading to a longer orbital period \citep{Hilditch}, which is the
opposite of what is observed, although the rate of such change is very low.

To explain the decrease in the orbital period, one has to assume that the
material carrying the angular momentum away must come from the vicinity of
the neutron star, since it is the neutron star's orbital motion that
represents the bulk of the angular momentum in the system. We can see two
mechanisms that could be responsible for the loss of angular momentum.
First, material of the stellar wind that is streaming by the neutron star
feels the gravitational pull of the moving neutron star. Only a fraction of
this material is eventually accreted onto the neutron star, and the larger
part is leaving the binary system and carrying some angular momentum away,
since the interaction with the neutron star changed its trajectory. Second,
before the matter is accreted onto the neutron star, it interacts with the
neutron star's magnetosphere (leading to spin-up and spin-down of the
neutron star, as will be discussed below). However, the interaction with
the magnetosphere may also lead to a magnetically driven outflow of
material \citep{Illarionov:1990p1675, Lovelace:1999p1780, klochkov2009},
again carrying angular momentum away. In addition, tidal coupling of the
rotational frequency of the optical star with the orbital frequency could
play some role, although estimates by \citet{Leahy:2008p358} and
\citet{Hilditch} suggest that, despite the high eccentricity, this is
probably not very efficient. The details of the mass transfer and angular
momentum loss in this system are not understood well, and more observations
are required to secure the rate of change in the orbital period. 

\subsection{Torque balance and magnetic moment of the neutron star}
The evolution of the spin frequency of the neutron star gives insight
into the interaction of the accretion flow with the neutron star. The
rotational dynamics is determined by the equation 
\begin{equation}
 I\frac{d\omega}{dt}=K_++K_-
\end{equation}
where $K_+$ and $K_-$ are the acceleration and deceleration torques, where
$\omega$ is the pulse frequency, and $I$ the momentum of inertia of the
neutron star. 

Two rapid spin-up episodes observed by~\cite{Koh:1997p138} indicate that a
long lived accretion disk may sometimes form in GX~301$-$2. Both episodes
are characterized by an increased source flux, which implies an increased
accretion rate. The infrequent occurrence of such episodes argues against
the hypothesis that they are triggered by tidal overflows at periastron
(see~\citealp{Layton}) and suggests that mass loss episodes of Wray~977 may
be responsible for them~\citep{Koh:1997p138}. As concluded by
\cite{Koh:1997p138}, the pulse period decrease in 1984-1992 can be
attributed entirely to similar spin-up episodes, while most of the time the
neutron star accretes from the wind, and no net change of the pulse period
is observed. It is therefore important to understand the torque balance in
this case. This is why we focus on wind-accretion models.

In the case of quasi-spherical accretion from a stellar wind the
accelerating torque can be expressed as \citep{Davies:1979p2881} 
\begin{equation}
 K_+=\dot{M}k_w{R_{A}}^2\Omega_\mathrm{orb}
 \label{eq:kw}
\end{equation}
where $\Omega_\mathrm{orb}$ is the orbital frequency, $\dot{M}$ the
accretion rate, $k_w$ dimensionless coefficient reflecting the efficiency
of the angular momentum transfer, and $R_A$ the accretion radius given by 
\begin{equation}
 R_{A}=\frac{2GM}{v_{rel}^2}\,,
\end{equation}
where $B$ is the field strength, $R,M$ are the neutron star radius and
mass, and $v_{rel}$ is the relative speed of the wind and the neutron star.
Hydrodynamical simulations~\citep{taam, Fryxell, ruffert:1992,
ruffert:1997} show that, due to the asymmetry of the accretion flow caused
by the orbital motion of the neutron star and due to fluctuations in the
speed and density of the local wind, short lived accretion disks may form.
The disk direction alternates between prograde and retrograde depending on
the local physical conditions. This causes the neutron star to alternate
between short spin-up (prograde disk) and spin-down (retrograde disk)
episodes, so a significant fraction of the angular momentum carried by the
wind is cancelled out. To account for this decrease in the efficiency of
angular momentum transfer, $k_w$ is introduced in Eq.~\ref{eq:kw}. Its
value is a controversial topic, but simulations predict that it is rather
small (absolute value less than $\sim1.2$ according
to~\citealt{Ho:1989p3226}). 

There are several models for the deceleration torque. According
to~\citet{Davies:1979p2881} and \citet{BisnovatyiKogan:1991p2029}, an
asymmetric magnetosphere of the accreting pulsar produces turbulent
viscosity in the nearby wind, which brakes the neutron star: 
\begin{equation}
 K_-=-\dot{M}\frac{\omega^2{R_m}^{7/2}}{4\sqrt{2GM}}
\label{eq:dav}
\end{equation}
where $R_m$ is the magnetospheric radius given by
\begin{equation}
 R_m={\left(\frac{B^2{R}^6}{2\dot{M}\sqrt{2GM}}\right)}^{2/7} 
\end{equation}

Instead, \cite{Illarionov:1990p1675} explain the spin down as a result of
the efficient angular momentum transfer from the rotating magnetosphere of
the accreting star to an outflowing stream of magnetized matter. This
outflow is said to be a result of the heating of the accreting matter by
hard X-ray emission from the pulsar through Compton scattering. The outflow
is focused within a certain solid angle owing the anisotropy of the pulsar
emission. It has a lower density than the surrounding accreting matter due
to the higher temperature, thus it is driven out by the buoyancy force.
Angular momentum gain from the accreting gas is balanced by angular
momentum loss via the outflow, enabling a spin-down of the neutron star
under certain circumstances. In this model, 
\begin{equation}
 K_-=-\dot{M}k\frac{\xi}{2\pi}{R_m}^2\omega\,,
\label{eq:k-}
\end{equation}
where $\xi\sim1$ is the solid angle of the stream outflow. The
dimensionless coefficient $k\sim2/3$ accounts for the orientation of the
outflow with respect to the magnetic field. It is worth noting, that the
equilibrium period obtained from the comparison of this torque with the
accelerating torque has the same value (aside from the numerical
coefficients) as second estimate by~\cite{BisnovatyiKogan:1991p2029}. The
braking torque there arises from the turbulent viscosity in the same way as
in Eq.~\ref{eq:dav}, but it is assumed, that the magnetosphere and
accreting matter corotate, and the velocity for the velocity dependent
viscosity coefficient is taken as the sound speed for the corresponding
temperature at the magnetosphere boundary (and not as relative velocity of
the wind and the magnetosphere). The angular momentum is then carried away
with the turbulent motions. As noted by \cite{BisnovatyiKogan:1991p2029},
the situation for real pulsars is probably somewhere in between these two
estimates. The braking torques in both models may also include undetermined
coefficients $\sim1$, which, can however, be incorporated into the
accelerating torque term. 

The torque balance, hence the rotational frequency derivative, depends on
$\dot{M}$, so one has to investigate this dependence to study the
rotational dynamics. Since the longest continuous pulse frequency
monitoring campaign (\hbox{$\sim10$\,yr) }for this source was carried out
with BATSE~\citep{Bildsten:1997p2328}, we used the data products available
at the
\href{ftp://heasarc.gsfc.nasa.gov/compton/data/batse/pulsar/histories/DISCLA_histories/gx301m2_psr_hist.fits}{CGRO }mission\footnote{\footnotesize{ftp:\-/\-/heasarc.gsfc.nasa.gov\-/compton\-/data\-/batse\-/pulsar\-/histories\-/DISCLA\_histories\-/gx301m2\_psr\_hist.fits}}.
The pulse frequency and pulse frequency derivative histories are provided
for the entire \textsl{BATSE} lifetime. Both were determined for a set of
$\sim4$\,d intervals using the phase connection technique assuming the
ephemeris by \cite{Koh:1997p138} for binary-motion corrections
(see~\citealp{Bildsten:1997p2328,Koh:1997p138} for details). The
corresponding \textsl{BATSE} pulsed flux in the 20--50\,keV energy range,
averaged over the interval is also provided. 

\begin{figure}[t]
\centering
\resizebox{\hsize}{!}{\includegraphics{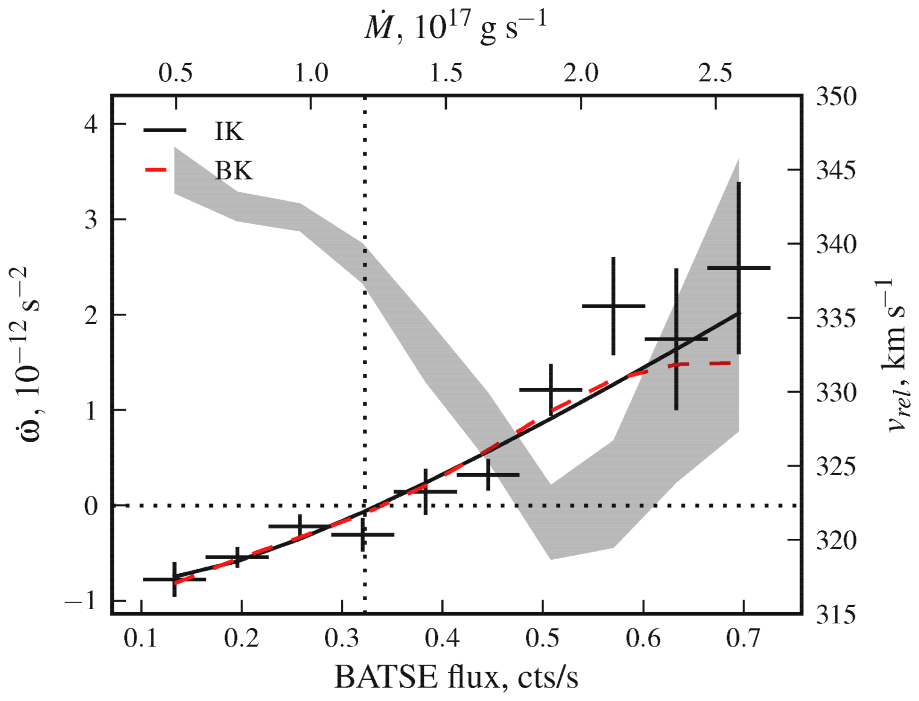}}
\caption{Angular frequency derivative correlation with flux. Flux
is \textsl{BATSE}-pulsed flux in the 20--40\,keV energy range. \textsl{BATSE
DISCLA} data provided by~\cite{Bildsten:1997p2328} on the \textsl{GGRO} mission
page are used to obtain the plot. Example of
fitting~\cite{Illarionov:1990p1675} (solid black) and
\cite{Davidson:1973p2909,Davies:1979p2881,BisnovatyiKogan:1991p2029}
(dashed red) models for an assumed distance of 3\,kpc is plotted. Top axis
shows the estimated accretion rate for this distance. The vertical line
indicates the mean flux during the observation. Shaded area represents
average relative velocity of the neutron star and wind for a given flux bin
(right scale).} 
\label{fig:flux_fdot}
\end{figure}

Contrary to the report by~\cite{Inam:2000p144}, a correlation between the
angular frequency derivative ($\dot{\omega}=2\pi\dot{\nu}$) and the flux
(see Fig.~\ref{fig:flux_fdot}) was found (Pearson correlation coefficient
0.96, null hypotheses probability 8$\times10^{-6}$). 

The discrepancy between our findings and the ones reported in
\cite{Inam:2000p144} lie in their method of estimating frequency
derivatives. With the \textsl{BATSE} data set that we used,
\cite{Inam:2000p144} estimate pulse frequency derivatives by grouping the
provided frequency values in intervals of $\sim30$\,d and averaging between
the left and right frequency derivatives calculated using these values for
each interval. This approach is incorrect because it assumes that the
frequency values provided by \cite{Koh:1997p138} alone characterize the
average pulse frequency during the corresponding observation time, while
the average pulse frequency also depends on the frequency derivative
included in the fit and on the observation length. For this approach to
work it is required to remove the frequency derivative in the fit for the
pulse arrival times in the raw \textsl{BATSE} data, which was not done by
\cite{Inam:2000p144}. There is also a second point to question in their
analysis. To obtain values of the first derivative, \cite{Inam:2000p144}
use frequency values on intervals of $\sim$30\,d, comparable to the orbital
period of the system. Both the pulse frequency and the flux are known to
change on much shorter time scales in GX~301$-$2. Averaging on such a long
interval smoothes out most of the flux and pulse frequency variations,
making it difficult to find the correlation between the two quantities. 

On the other hand, we used $\dot{\omega}$ and flux values directly measured
for each observation with phase connection. The points in
Fig.~\ref{fig:flux_fdot} were obtained by averaging provided frequency
derivative values of points with flux in a given range. The standard error
was used as an uncertainty estimate. We excluded both spin-up episodes
observed by~\cite{Koh:1997p138} (i.e. MJD~48440--48463 and
MJD~49230$-$49245) and intervals where pulsations were not detected
reliably (see \citealt{Koh:1997p138}) from the further analysis. 

To investigate the accretion models and compare them to the data, we need
to express the accretion rate as a function of the count rate, not a
trivial task. The conversion depends on the distance, radiative efficiency
of accretion, and beaming factor. We assumed that the mean source flux
derived from the spectra obtained with the \textsl{INTEGRAL} pointed
observations corresponds to the mean \textsl{BATSE} count-rate at the same
orbital phase. Then we assumed a conversion factor of
$10^{37}$\,erg\,s$^{-1}\simeq10^{17}$\,g\,s$^{-1}$ which corresponds to the
radiative efficiency of accretion $\sim10$\%,
$L_\mathrm{x}\sim0.1\dot{M}\rm{c}^2$) to estimate the accretion rate. The
distance to the source is uncertain so the derived value should account for
the spread of the estimates (1.4-5.3\,kpc). 
Apparently, $K_+$, hence the torque balance, depends significantly on the
efficiency of angular momentum transfer $k_w$ and on the relative
velocity of the neutron star and the wind. 

\begin{figure}[t]
	\centering
		\includegraphics[scale=0.5]{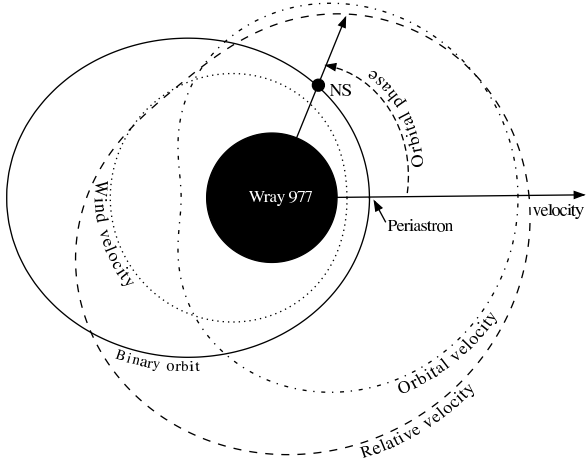}
 \caption{Sketch of the GX~301$-$2 system. 
The relative speed of the neutron star and wind, the orbital velocity of the
neutron star, and the intrinsic velocity of the wind close to the neutron
star are plotted as a function of the orbital phase. Velocities are
normalized to maximal relative velocity $\sim380$\,km\,s$^{-1}$, and distances
to the periastron distance $\sim1.75\times10^{13}$\,cm.}
\label{fig:diagram} 
\end{figure}

The orbit of GX~301$-$2 is eccentric, so the orbital speed of the neutron
star changes significantly along the orbit. The wind, on the other hand, is
also accelerated from the sound speed at the surface of Wray~977
($\sim10$\,km\,s$^{-1}$) to a terminal velocity of 300--400\,km\,s$^{-1}$
at infinity~\citep{castor}: 
\begin{equation}
	v_w(r)=v_0+(v_\infty-v_0)(1-R_*/r)^\beta
\end{equation}
where $v_0$ is the velocity at the surface of the star close to the sound
speed, $v_\infty$ the terminal wind speed, and $\beta\sim1$ for O-type
stars. Radial and tangential components of the neutron star as function of
orbital phase $\theta$ are 
\begin{equation}
	v_r=\sqrt{\frac{\mu}{p}}e\sin{\theta},\,\,\,\,v_t=\sqrt{\frac{\mu}{p}}(1+e\cos{\theta})\\
\end{equation}
where $\mu=G(\mathrm{M}_{\mathrm{opt}}+\mathrm{M}_{\mathrm{NS}})$,
$p=a(1-e^2)$, $e$ is the eccentricity, and $a\sim1.2\times10^{13}$\,cm is
the semi-major axis for $i=66^\circ$. It turns out that, while both the
orbital speed of the neutron star and the wind speed are strong functions
of the orbital phase, the relative velocity varies only by a factor of 2
(see Fig.~\ref{fig:diagram}). 
Each flux bin in Fig.~\ref{fig:flux_fdot} contains measurements
performed at different orbital phases. We then calculated an average
relative speed for each flux bin to properly estimate $R_A$. It turns out,
however, that the relative speed varies only within 5\% and consequently
$R_A$ does not changes significantly (see Fig.~\ref{fig:flux_fdot}). 

\begin{figure}[t]
\centering
 \includegraphics{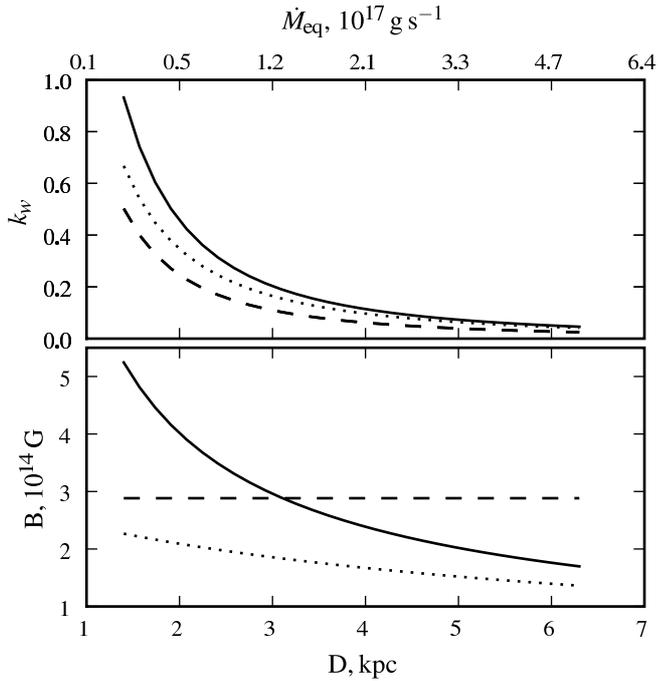}
\caption{Results of fitting frequency derivative -- flux correlation (see
Fig.~\ref{fig:flux_fdot}) with~\cite{Illarionov:1990p1675} (solid),
\cite{Davidson:1973p2909}, \cite{Davies:1979p2881},
\cite{BisnovatyiKogan:1991p2029} (dashed) and a model with both braking
torques in place (dotted) depending on assumed distance, hence mean
accretion rate. The required magnetic field strength B (bottom pane)
depends on the efficiency of angular momentum transfer $k_w$ (top pane),
which is constrained by the fit.}
\label{fig:kwb}
\end{figure}

The absolute value of $k_w$ cannot be arbitrarily small, otherwise the
source will not be able to spin up, while this is clearly the case
when the flux exceeds a certain value. The value of $k_w$ can, therefore, be
estimated using the observed frequency derivative over accretion rate
dependence. We only measure flux, so the accretion rate and $k_w$ are
parametrized by the assumed distance. For each distance in the range of
published estimates from 1.4 to 5.3\,kpc, we calculated the accretion rate
and estimated $k_w$ and $B$ as free parameters of models defined by
Eqs.~\ref{eq:kw}--\ref{eq:k-} where we assumed M=1.4$M_\odot$,
R=$10^6$\,cm, I=1.4$\times10^{45}$\,g\,cm$^2$,
$v_\infty$=300\,km\,s$^{-1}$, k=2/3, and $\xi$=0.87. The estimated values
are presented in Fig.~\ref{fig:kwb}. The $k_w$ range is in line with
estimates obtained by~\cite{Ho:1989p3226} and with later claims that the
average amount of angular momentum transferred from the wind to the neutron
star is relatively small~\citep{ruffert:1992,ruffert:1997}. It is
likely that the mechanisms to generate the braking torques assumed in the
models may act simultaneously, so we attempted to find the magnetic field
required in this case by including both torques. The required field
strength, however, is not significantly reduced and still exceeds
$10^{14}$\,G (see Fig.~\ref{fig:kwb}). 

It is important to emphasize that the frequency derivative and therefore
torque affecting the neutron star are consistent with zero for the average
source flux. This means that, for the average conditions during the
observations, the period is close to a so-called equilibrium period (i.e.
when the torques are balanced). This is also in line with the long-term
pulse period evolution (see Fig.~\ref{fig:perhist}). The knowledge of the
equilibrium period allows estimation of the magnetic field for the average
luminosity even without knowing the exact dependency of torque with
luminosity. For example, for the~\cite{Davidson:1973p2909},
\cite{Davies:1979p2881}, \cite{BisnovatyiKogan:1991p2029} model in the case
of torque equivalence, the field strength may be expressed as 
\begin{eqnarray}
B \approx 3\times10^{14}\,\mathrm{G}\left(\frac{k_w}{0.25}\right)^{1/2}\left(\frac{\dot{M}_\mathrm{eq}}{10^{17}\,\mathrm{g/s}}\right)^{1/2}\left(\frac{v_{rel}}{400\,\mathrm{km/s}}\right)^{-2} \\ \nonumber
\times \left(\frac{P}{680\,\mathrm{s}}\right)\left(\frac{P_\mathrm{orb}}{41.5\,\mathrm{d}}\right)^{-1/2}\left(\frac{M}{1.4M_\odot}\right)^{3/2}\left(\frac{R}{10^6\,\mathrm{cm}}\right)^{-3}\nonumber.
\end{eqnarray}
The equivalent equation for the~\cite{Illarionov:1990p1675} model is
\begin{eqnarray}
B \approx 2\times10^{14}\,\mathrm{G}\left(\frac{k_w}{0.25}\right)^{7/8}\left(\frac{k}{2/3}\right)^{-7/8}\left(\frac{\xi}{0.87}\right)^{-7/8}\left(\frac{\dot{M}_\mathrm{eq}}{10^{17}\,\mathrm{g/s}}\right)^{1/2}\\ \nonumber
\left(\frac{v_{rel}}{400\,\mathrm{km/s}}\right)^{-7/2} \left(\frac{P}{680\,\mathrm{s}}\right)^{7/8}\left(\frac{P_\mathrm{orb}}{41.5\,\mathrm{d}}\right)^{-7/8}\left(\frac{M}{1.4M_\odot}\right)^{2}\left(\frac{R}{10^6\,\mathrm{cm}}\right)^{-3}\nonumber.
\end{eqnarray}
The strength of the magnetic field calculated under the
assumption of an equilibrium period using the models for systems accreting
from the persistent disk~\citep{Lovelace:1999p1780,Ghosh:1979p1676} is even
\hbox{stronger~($\sim10^{15}$\,G).} 
\subsection{Implications of the strong magnetic field.}
As shown above, a very strong magnetic field is required to explain
the long pulse period of the GX~301$-$2. Other authors~\citep{Li:1999p3188}
argue that a long-period pulsating source might be a relic of magnetar
evolutionary phase, when the field is strong enough to sufficiently spin
the neutron star down, and demonstrate the feasibility of this scenario for
\hbox{2S~0114+650}. However, this does not apply to GX~301$-$2, since the
source is in torque equilibrium and therefore the field has to be very
strong presently, unless there are some unidentified braking torques. The
observed CRSF energy, on the other hand, corresponds to a field of
$B\sim4\times10^{12}$\,G, which contradicts our previous
conclusion. 

\begin{figure}[t]
\centering
\includegraphics{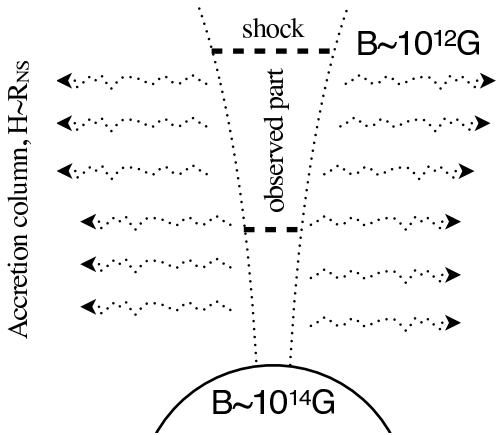}
\caption{Sketch of a radiation-dominated accretion column.
Temperature and magnetic field strength increase towards the neutron star surface.}
\label{fig:sketch}
\end{figure}

We suggest that it may be resolved if the line-forming region is situated
far above the neutron star's surface (i.e. in the accretion column). To
reconcile a surface field of $\sim10^{14}$\,G with the one derived from the
observed CRSF energy, one must assume that the CRSF is formed at height 
$H\sim R_\mathrm{NS}(({B_\mathrm{surf}/B_\mathrm{CRSF}})^{1/3}-1)\sim2-3\,R_\mathrm{NS}$
(see Fig.~\ref{fig:sketch}). 

The accretion column rises owing to the radiation pressure when the flux
from the hotspot on the neutron star surface becomes comparable to a
critical flux (local Eddington luminosity). The column height increases
with the accretion rate to allow the excess energy to radiate away from the
side surface. The observed color temperature 4--5\,keV of the
\hbox{GX~301$-$2} spectrum suggests that the accretion column is likely to
form in this source. Indeed, the effective critical temperature, which
corresponds to the critical flux, is $T_{\rm Edd} \sim$2\,keV for standard
neutron star parameters ($M=1.4\,M_\odot$, $R$=10$^6$\,cm) and solar
composition of the accreting matter. The observed spectrum is expected to
be close to the diluted Plank spectrum $B_E$: $F_E \approx B_E(T_c)/f_c^4$,
with a color temperature $T_c = f_c T_{\rm{eff}}$ and a hardness factor
\hbox{$f_c\sim$1.5--2} because of the Compton
scattering~\citep{pavlov1991}. This qualitative picture is similar in the
case of Compton scattering in the strong magnetic
field~\citep{Lyubarskii:1986p3028}, so the measured color temperature
probably corresponds to a critical effective temperature at the neutron
star surface (or at a certain height above the surface, but in this case
the temperature at the surface is expected to be even higher, so the
accretion column forms anyway). 

It is possible to estimate the height of the accretion column, and it turns
out to be compatible with the requirement mentioned above:
$H\sim2-3\,R_\mathrm{NS}$. Indeed, the accretion column base radius can be
estimated from the neutron star magnetic moment $r \approx R_\mathrm{NS}
(R_\mathrm{NS}/R_{\rm H})^{1/2}$~\citep{Lipunov:1992p2320}. The
magnetospheric radius is $\approx(3-30)\times10^8$\,cm for a magnetic field
in the range $10^{12}-10^{14}$\,G. The corresponding radius of the column
base is \hbox{200--500\,m}. The accretion column height may then be
estimated using cylindrical geometry, and the critical effective temperature
from the observed luminosity $\displaystyle L \approx 10^{37}$ erg s$^{-1}$
$\approx 2~ \sigma_{\rm SB} T_{\rm Edd}^4~ 2\pi r$. This
simple estimate gives $H \approx$ 8--20\,km (for B=$10^{12}-10^{14}$\,G and
$\dot{M}=1.2\times10^{17}$\,g\,s$^{-1}$). More elaborate calculations by
\cite{Basko:1976p1538}, and \cite{Lyubarsky1988} give similar results (depending on the
assumed accretion column geometry). 

As shown by~\citet[see Fig.~4 and accompanying discussion]{Basko:1976p1538}, 
the amount of energy released by a unit of
height of the accretion column is almost constant, so a significant part of
emission comes from the outer parts of it. The contribution of the outer
parts is especially important because the inner parts of the column are
more easily obscured by the neutron star. It is natural to assume that
X-ray emission and cyclotron line formation regions coincide, so the whole
column contributes to the formation of the CRSF. The magnetic field 
strength and the other physical properties depend on height, so a
mix of cyclotron lines coming from regions with different physical
properties are observed~\citep{Nishimura:2008p2969}. The contribution
of the outer parts of the column to the cyclotron line formation is
especially important because the field there is weaker, so the
line is observed at lower energies, which makes it easier to detect.

The formation of the pulse profile and the spectrum of a tall accretion
column has not yet been understood and is beyond the scope of this work,
although some characteristic features may be reckoned. The color
temperature in the column increases towards the neutron star surface. The
inner and hotter parts of the column are more likely to be obscured by the
neutron star, so the pulse fraction increases with temperature (hence
energy). For certain pulse phases, the inner parts of the column are
obscured by the neutron star, and we can see only the outer, relatively
cool parts, while a larger part of the column is seen at other phases. The
pulse fraction is also expected to decrease with the increase in the column
height and therefore with the source luminosity, which is indeed the case
for GX~301$-$2~\citep{lutovinov}. 

The magnetic field strength increases towards the neutron star, so one can
expect the centroid energy of the CRSF to be anti-correlated with total
column height (hence luminosity) and correlated with the effective color
temperature (or the cutoff energy, which is believed to be proportional to
temperature). The latter is indeed the case for \hbox{GX~301$-$2}, as
reported by~\citet[see~Table~3]{Kreykenbohm:2004p155}, while the
correlation of the line energy with the luminosity is not confirmed because
the cyclotron line was never detected outside of the pre-periastron flare. 

When the vertical span of the observed part of the column is short, the
line width is also expected to be smaller. In contrast, the line becomes
wider when a larger part of the column is observed as the magnetic field
increases by an order of magnitude from the column top to the bottom. This
could explain the ``line width - line energy'' correlation reported by
\citet{Kreykenbohm:2004p155} for GX~301$-2$ and also by
\citet{Coburn:2002p158} for several other sources. As was concluded by
\citet{Coburn:2002p158}, the observed correlation is independent of the
spectral model used and is not affected by selection effects. It can
therefore be considered as an intrinsic correlation between CRSF
parameters. The proposed scenario with a tall accretion column explains it
and complements the explanation of this correlation by
\citet{Coburn:2002p158}, who attribute it to a change in the viewing angle
with respect to the magnetic field (see also
\citealt{1992herm.book.....M}). 

The cyclotron line formation in a tall column with temperature gradient was
investigated in detail by~\cite{Nishimura:2008p2969} with similar
conclusions. He considered a surface field of $\sim10^{12}$\,G and about
an order of magnitude less at the top of the accretion column. Both the
height of the column and the range of the physical parameters within it are
expected to be greater for a stronger surface field, but the results should
be similar. Detailed modeling of the accretion column which accounts for
light-bending and beaming is, however, required to clarify the expected
shape, and pulse-phase dependence of the spectrum. The pulse profile shape
and phase dependence of the spectrum in
GX~301$-$2~\citep{Kreykenbohm:2004p155} is similar to other luminous
slow-pulsating sources like~\hbox{Vela~X-1}~\citep{vela} and
\hbox{4U~1538-52} \citep{4u1538}. Common spectral features include a
correlation of CRSF energy with the width and with the cutoff energy 
(which characterizes the continuum temperature) in phase-resolved spectra,
so the discussion above may also be relevant for these objects. 
 
It is worth noting that the mass of the optical counterpart is very high,
so one can expect that the second supernova explosion in the system may
take place before the magnetic field decays significantly. Such an
explosion will most likely disrupt the system and leave an isolated neutron
star with a magnetar-like magnetic field and the long pulse period.
The observational appearance of such an object is unclear. The strong
magnetic field might power a magnetar-like emission, however, known
magnetars (i.e. soft gamma-repeaters and anomalous X-ray pulsars) have much
shorter pulse periods. 
\section{Summary and conclusions} In this work we studied the timing
and spectral properties of \hbox{GX~301$-$2} using the archival data of
\textsl{INTEGRAL} and data products of \textsl{CGRO BATSE} and
\textsl{RXTE ASM}. An orbital-period's secular change was detected and the
pulse period history since May~2005 determined. This shows a steady
spin-down trend. The apparent rate of decay of the orbital period is about an
order of magnitude higher than for other known sources. We argue that
this is probably caused by angular momentum loss by material expelled from the
vicinity of the neutron star to the outside world. However, we cannot at this time
exclude some contribution from a possible apsidal motion to the observational
appearance. 

Results of our spectral analysis are consistent with previous works,
although we find that the spectrum is described well not only with
phenomenological models, but also with a saturated comptonization model. 

We discussed a possible scenario to explain the long pulse period and
long spin-down trends observed despite steady accretion of matter and
angular momentum onto the neutron star. We studied the balance of the
torques affecting the neutron star using \textsl{BATSE/DISCLA} data by
\cite{Bildsten:1997p2328} and find that the rotational frequency
derivative is correlated with the flux. We also find that the frequency
derivative is zero for the average count rate, which is a signature that
the observed pulse period reflects torque equilibrium during the
observations' time span. The scenario invoked by~\cite{Li:1999p3188} to
explain the long period of \hbox{2S~0114+650} cannot therefore be applied
to \hbox{GX~301$-$2} since the observed pulse period is close to equilibrium.
We investigated several published torque models to constrain the magnetic
field strength and found that all of them require the field to be
$\ga10^{14}$\,G. The magnetic field strength derived from the observed
CRSF energy turns out to be $\sim4\times10^{12}$\,G, i.e. at least an order
of magnitude less than from the timing. We argue that this can be
explained if the line-forming region resides high up in the accretion
column. We show that the accretion column as high as $\sim$ 10 -- 20\,km is
expected to form in \hbox{GX~301$-$2} in the framework of
the~\cite{Basko:1976p1538} model and that it is sufficient to reconcile the
very strong field at the surface with the observed cyclotron line energy.
Following the scenario proposed by~\cite{Nishimura:2008p2969}, we conclude
that these correlations may be explained qualitatively by a simultaneous
change of height and vertical span of the observed region with pulse phase
if the line-forming region resides in a tall accretion column with a
temperature gradient. The quantitive description of the spectrum with a
model such as the one developed by \cite{Nishimura:2008p2969} and detailed pulse
profile formation modeling is, however, required to confirm this
scenario. 

An alternative scenario is that the long pulse period is explained by the
presence of some unidentified braking torque, which is less dependent on
magnetic field strength, and the CRSF pulse phase variability is attributed,
as concluded by~\cite{Kreykenbohm:2004p155}, to multipole
field components. 

\begin{acknowledgements}
We sincerely wish to thank N.~Shakura, K.~Postnov, and the anonymous
referee for the useful comments and discussions, which helped to improve
the paper. V.D., D.K., and C.F thank the DFG for financial support (grants
DLR~50~OR~0702 and DLR~50~OG~0601). VS thanks the DFG for financial support
(grant SFB/Transregio~7 ``Gravitational Wave Astronomy''), and for partial support the RBRF(grant
\hbox{09-02-97013-p-povolzh'e-a}). We also acknowledge the support of the
International Space Science Institute (Bern). This research is based on
observations with INTEGRAL, an ESA project with the instruments and science
data center funded by ESA member states (especially the PI countries:
Denmark, France, Germany, Italy, Switzerland, Spain), the Czech Republic, and
Poland, with the participation of Russia and the USA. 
\end{acknowledgements}
\bibliography{12951}

\end{document}